\documentclass[aps,prc,preprint,groupedaddress,showpacs]{revtex4}
\usepackage{epsfig}
\usepackage{graphicx}
\usepackage{amsmath}
\usepackage{amsfonts,amssymb}
\usepackage{color}
\usepackage{braket}
\usepackage{slashed}
\usepackage{url}

\DeclareRobustCommand{\URL}{\url}

  \usepackage{amsthm}
\usepackage{bm}
\usepackage{verbatim}
\usepackage{float}

\begin{document}

%\preprint{}

\title{Condensates in the refined Gribov-Zwanziger scenario}


\author{Marco Frasca}
\email[e-mail:]{marcofrasca@mclink.it}
%\thanks{}
\affiliation{Via Erasmo Gattamelata, 3 \\
             00176 Roma (Italy)}


\date{\today}

\begin{abstract}
Recent lattice computations showed how the approach dubbed ``refined Gribov-Zwanziger scenario'' is in very good agreement with data and the gluon propagator fits them very well. This propagator can be described as a finite sum of Yukawa propagators and can be obtained introducing some condensates representing the contribution of the vacuum of Yang-Mills theory. The values of the condensates are arbitrary and are obtained through lattice data. This kind of structure of the propagator is in agreement with the one we obtained using a different analytical approach but with the substantial difference that here all the physical parameters are properly fixed. We show that our approach can properly fix all the values of the condensates in the refined Gribov-Zwanziger scenario giving a complete validation to both techniques. This will provide an interesting view to the gluon propagator and a set of values to be experimentally determined. 
\end{abstract}

\pacs{12.39.Fe, 11.30.Rd, 12.38.Mh}
%\keywords{}

\maketitle


\section{Introduction}
%\label{}
%%%%%%%%%%%%

In recent years, lattice studies of the two-point functions made clear their structure \cite{Bogolubsky:2007ud,Cucchieri:2007md,Oliveira:2007px}. These results have shown how a previous view \cite{von Smekal:1997is,von Smekal:1997vx} about gluon and ghost propagators for a pure Yang-Mills theory was diverging in giving an explanation to their behavior. The question that was put forward by Gribov \cite{Gribov:1977wm} and then improved by Zwanziger \cite{Zwanziger:1989mf} showed that the gluon propagator should go to zero as momenta go to zero and the ghost propagator should go to infinity faster than a free propagator. These conditions granted that positivity was maximally violated and the theory is confining. The new paradigm emerging from lattice computations showed instead a rather different situation: Gluon propagator reached a finite non-null value at zero momenta and the ghost propagator was indeed behaving as that of a free particle. This situation is in good agreement with the idea of a massive gluon (a mass gap) for Yang-Mills theory but appeared to conflict with the initial ideas about confinement.

From a theoretical standpoint, the pioneering work of Cornwall in the '80 was claiming for the idea of a non-null propagator at zero momenta \cite{Cornwall:1981zr}. At the start of this century this idea was revamping \cite{Aguilar:2004sw,Boucaud:2005ce} with numerical analysis of Dyson-Schwinger equations and the theoretical work that followed \cite{Aguilar:2006gr} to support this conclusion.

Between different theoretical approaches that are emerging in order to give an understanding to the unexpected behavior of the Green functions of Yang-Mills theory, the idea that condensates are playing some relevant role in the vacuum of the theory was initially put forward \cite{Dudal:2003by}. This scenario has been recently improved in such a way to completely improve Gribov-Zwanziger scenario in a ``Refined Gribov-Zwanziger'' scenario (RGZ for the following) \cite{Dudal:2008sp,Dudal:2008rm,Dudal:2011gd} that reached a substantial agreement with lattice computations \cite{Dudal:2010tf,Cucchieri:2011ig,Cucchieri:2012cb}. These comparisons with lattice data are a proof that condensates, that are generally postulated in this approach, indeed exist and this way to analyze the low-energy behavior of Yang-Mills theory is sound.

The structure of the gluon propagator that emerges from these studies has the form of a sum of free propagators of massive particles. This idea was firstly put forward by Migdal in the '70 \cite{Migdal:1977nu}. Recent works by Bochicchio for large-N Yang-Mills theory give strong support to Migdal's conclusion \cite{Bochicchio:2011en,Bochicchio:2011ks,Bochicchio:2012bj}.

In view of these relevant conclusions, it appears really important to derive from first principles the values of the condensates and make the analysis complete from numerical to a theoretical one. In order to accomplish this task, we will use a technique devised by us that has the advantage to have all the parameters properly derived from the theory and to get the proper structure of the gluon propagator as a sum of free massive propagators \cite{Frasca:2007uz,Frasca:2009yp,Frasca:2011bd}. In this way, we will see this as a validation to both these approaches being them each other consistent. The idea behind this technique is to take for granted, at some stage, that the emerging data from lattice about the running coupling of the theory are just displaying a trivial infrared fixed point for a pure Yang-Mills theory \cite{Bogolubsky:2009dc}. Then, by proving the existence of a class of classical solutions to Yang-Mills theory, this behavior is shown to emerge and, combined with a Gaussian form of the generating functional in the in the infrared limit, we can build up a consistent framework to compare with RGZ scenario.

The paper is so structured. In sec.\ref{sec1} we discuss the gluon propagator as given in RGZ scenario and derive it from instanton solutions obtained from the classical equations of motion of the Yang-Mills theory, starting from lattice data showing the running coupling to go to zero lowering momenta. In sec.\ref{sec2} we derive the condensates of RGZ scenario in a closed form proving also that the agreement is consistent with a sum of two Yukawa propagators as also is emerging in RGZ studies and fits with lattice data. Finally, in sec.\ref{sec3} conclusions are given.

%%%%%%%%%%%%
\section{Gluon propagator}
\label{sec1}
%%%%%%%%%%%%

\subsection{Refined Gribov-Zwanziger scenario}

The initial formulation of the Gribov-Zwanziger scenario was given in 1989 \cite{Zwanziger:1989mf}. A completely renormalizable Lagrangian of Yang-Mills theory was formulated that agrees at all orders with Gribov prescription to restrict the path integral to the first Gribov region. This was achieved with the introduction of a couple of bosonic fields $(\varphi_\mu,\bar\varphi_\mu)$ and a couple of fermionic ones $(\omega_\mu,\bar\omega_\mu)$. This Lagrangian provides the following prescriptions for the gluon propagator in the Landau gauge in the infrared limit for SU(N)
\begin{equation}
   D_{\mu\nu}^{ab}(p^2)=\delta_{ab}\left(\eta_{\mu\nu}-\frac{p_\mu p_\nu}{p^2}\right)D(p^2)
   =\delta_{ab}\left(\eta_{\mu\nu}-\frac{p_\mu p_\nu}{p^2}\right)\frac{p^2}{p^4+2Ng^2\gamma^4}
\end{equation}
with $\gamma$ is a constant named Gribov parameter. This propagator goes to zero lowering momenta and so is in disagreement with lattice computations. Similarly, this scenario gets a ghost propagator infrared enhanced as it goes to infinity faster than the free propagator in the same limit, again in disagreement with lattice data. But this scenario can be reconciled with numerical data when one notes that the effective potential in the Gribov-Zwanziger theory permits the presence of condensates \cite{Dudal:2008sp,Dudal:2008rm,Dudal:2011gd}. Then, the most general gluon propagator can be written down in the Landau gauge as \citep{Dudal:2011gd} as
\begin{equation}
D(p^2) \; = \;\frac{p^4 + 2 M^2 p^2+ M^4 - \rho \rho^\dagger}{
                    p^6 + p^4 \left(m^2 + 2 M^2 \right) + p^2 \left(2 m^2 M^2 + M^4 + \lambda^4 - \rho \rho^\dagger \right)
                  + m^2 \left( M^4 - \rho \rho^\dagger \right)
                  + M^2 \lambda^4 - \frac{\lambda^4}{2}
                  \left( \rho + \rho^\dagger \right)} \; ,
\label{eq:Drho}
\end{equation}
where the condensates are given by
\begin{align}
\braket{A_\mu^a A_\mu^a}                                & \to -m^2  &
\Braket{\overline{\varphi}^a_i \varphi^a_{i}}           & \to M^2  &
\Braket{\varphi^a_i \varphi^a_{i}}                      & \to \rho &
\Braket{\overline{\varphi}^a_i \overline \varphi^a_{i}} & \to \rho^\dagger \; ,
\label{eq:condensates}
\end{align}
and $\lambda^4$ is related to the Gribov parameter $\gamma$ through
$\lambda^4 = 2 g^2 N \gamma^4$. We note that, when $\Braket{\varphi^a_i \varphi^a_{i}}=\Braket{\overline{\varphi}^a_i \overline \varphi^a_{i}}$ then $\rho=\rho^\dagger$, and the propagator reduces to
\begin{equation}\label{prop}
D(p^2) \; = \; \frac{p^2 + M^2 + \rho_1}{p^4 + p^2 \left( M^2 + m^2 +
      \rho_1 \right) + m^2 \left( M^2 + \rho_1 \right) +  \lambda^4} \;.
\end{equation}
These propagators have the nice property to have $D(0)\ne 0$ and so, they agree with lattice data. One of the most interesting properties of these propagators is that they can be rewritten as a sums of Yukawa propagators. So, for eq.(\ref{eq:Drho}) one has
\begin{equation}\label{gluonpropsimp}
D(p^2) \; = \; \frac{\alpha}{p^2+\omega_1^2} \,+\, \frac{\beta}{p^2+\omega_2^2}
\,+\, \frac{\gamma}{p^2+\omega_3^2} \;\,.
\end{equation}
and for eq.(\ref{prop}) is
\begin{equation}\label{4D3}
D(p^2) \; = \; \frac{\alpha_+}{p^2+\omega_{+}^2} \,+\,
\frac{\alpha_-}{p^2+\omega_{-}^2} \;.
\end{equation}
This kind of propagators are in agreement with the scenario recently emerged from lattice computations. We will discuss this in the following sections showing how all this agrees with the current understanding of the Yang-Mills scenario on the lattice.

\subsection{Gluon propagator at the trivial infrared fixed point}

Computations on the lattice for a pure Yang-Mills theory showed that the running coupling manifests a peculiar behavior: The infrared limit is seen to reach zero lowering momenta \cite{Bogolubsky:2009dc}.

\begin{figure}[H]
\begin{center}
\includegraphics{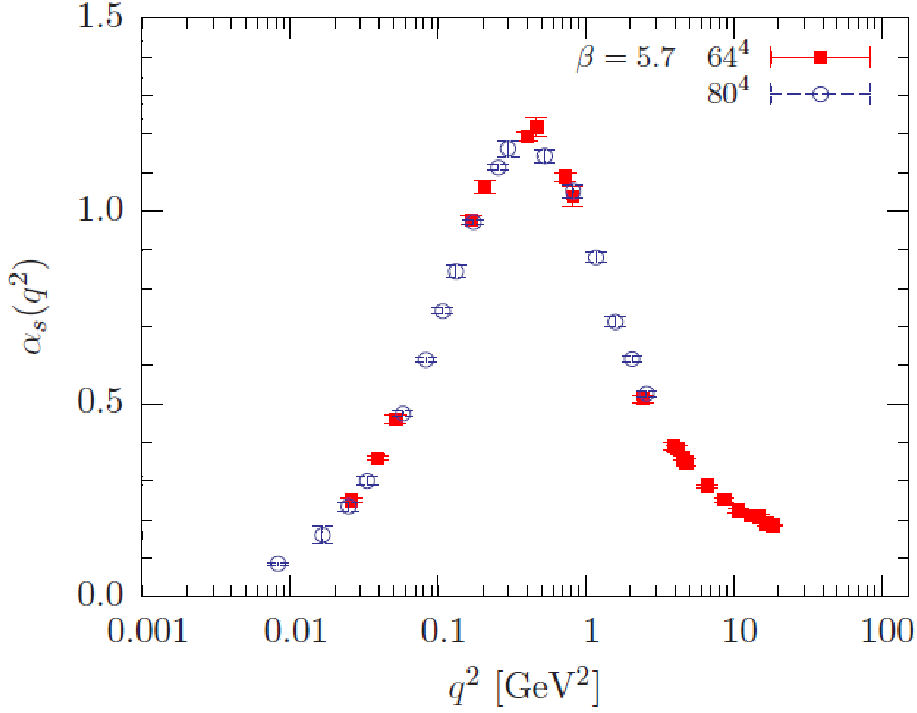}
\caption{Running coupling for $64^4$ and $80^4$ at $\beta=5.7$ taken from Ref.\cite{Bogolubsky:2009dc} (\URL{http://www.sciencedirect.com/science/journal/03702693}, courtesy of Andre Sternbeck).\label{fig:stern}}
\end{center}
\end{figure}

From fig.\ref{fig:stern} it is easily realized that the coupling reaches  trivial fixed points at higher and lower energies reaching a maximum in the intermediate regime. It must be emphasized that this is only true for a pure Yang-Mills theory. In QCD, the presence of quarks makes this fixed point non trivial. A trivial fixed point implies a Gaussian leading order functional for the partition function and we could write
\begin{equation}
   Z_{YM}[j]\sim Z[0]e^{i\int d^4xd^4yj^{\mu a}(x)D^{ab}_{\mu\nu}(x-y)j^{\nu b}(y)}.
\end{equation}
This kind of generating functional can be understood if we assume, as done in '80 \cite{Cahill:1985mh}, that at low energies the functional of a Yang-Mills theory admits a series expansion in powers of the currents. Lattice data confirms {\sl a posteriori} this hypothesis. Now we will show that exists a set of solutions for the inhomogeneous equations of motion of the Yang-Mills theory that admits a propagator of a free theory. Before to accomplish this, we note that, for gauge theories, one can devise a reformulation of Wightman axioms as done by Strocchi \cite{Strocchi:1993wg} that provides a K\"allen-Lehman representation with a non-positive spectral function. If this is the case, the most general form of propagator for a free theory takes the from
\begin{equation}ù
\label{eq:propf}
   D(p^2)=\sum_{n=0}^\infty\frac{Z_n}{p^2-m_n^2+i\epsilon}
\end{equation}
being $m_n$ an internal spectrum of the free particle. Indeed, to put in Gaussian form the generating functional, we have to show that exists at least a gauge choice for classical Yang-Mills equations having as a solution eq.(\ref{eq:propf}). So, following a standard notation \cite{Peskin:1995ev}, we take for the generators
\begin{equation}
  [t^a,t^b]=if^{abc}t^c
\end{equation}
being $f^{abc}$ the structure constants of the gauge group. Then, for the covariant derivative we have
\begin{equation}
   D_\mu=\partial_\mu-igt^aA_\mu^a
\end{equation}
with $g$ the coupling constant and $A_\mu^a$ the potential. This gives
\begin{equation}
\label{eq:F}
   F_{\mu\nu}^a=\partial_\mu A^a_\nu-\partial_\nu A^a_\mu+gf^{abc}A^b_\mu A^c_\nu.
\end{equation}
So, the equation of motion are
\begin{equation}
   \partial^\mu F_{\mu\nu}^a+gf^{abc}A^{\mu b}F^c_{\mu\nu}=-j^a_\nu
\end{equation}
being $j^a_\nu$ arbitrary currents. To these equations we can add a gauge fixing term $\partial_\nu (\partial\cdot A^a)/\xi$. Now, we can prove that exist a solution of these equations with a propagator given by eq.(\ref{eq:propf}). Taking into account eq.(\ref{eq:F}) one has
\begin{equation}
\label{eq:y-m}
   \partial^\mu\partial_\mu A^a_\nu-\left(1-\frac{1}{\xi}\right)\partial_\nu (\partial\cdot A^a)+gf^{abc}\partial^\mu(A^b_\mu A^c_\nu)
   +gf^{abc}A^{\mu b}(\partial_\mu A^c_\nu-\partial_\nu A^c_\mu)+g^2f^{abc}f^{cde}A^{\mu b}A^d_\mu A^e_\nu
   =-j^a_\nu.
\end{equation}
Our aim is to look for a set of instanton solutions \cite{Schafer:1996wv} that can be cast into the form \cite{Weiss:1980rj}
\begin{equation}
   A^a_\mu=\eta_\mu^a\phi(x)
\end{equation}
being $\eta_\mu^a$ some coefficients that can carry some dependence on momenta. Then, by a direct substitution into eq.(\ref{eq:y-m}) we get
\begin{equation}
\label{eq:phi}
   \eta^a_\nu\partial^\mu\partial_\mu\phi-\left(1-\frac{1}{\xi}\right)\partial_\nu (\eta^a\cdot\partial\phi)+gf^{abc}\eta^b_\mu\eta^c_\nu\partial^\mu(\phi^2)
   +gf^{abc}\eta^b_\mu\phi(\eta^c_\nu\partial_\mu\phi-\eta_\mu^c\partial_\nu\phi)+g^2f^{abc}f^{cde}\eta^{\mu b}\eta^d_\mu\eta^e_\nu\phi^3
   =-j^a_\nu.
\end{equation}
By multiplying by $\eta^{\nu a}$ both sides we arrive at
\begin{equation}
\label{eq:phi2}
   \eta^{\nu a}\eta^a_\nu\partial^\mu\partial_\mu\phi-\left(1-\frac{1}{\xi}\right)\eta^a\cdot\partial (\eta^a\cdot\partial\phi)
   +g^2\eta^{\nu a}f^{abc}f^{cde}\eta^{\mu b}\eta^d_\mu\eta^e_\nu\phi^3
   =-j.
\end{equation}
Now, we note that $\eta^{\nu a}\eta^a_\nu=N^2-1$ (one can see this immediately by considering the SU(2) case) and the equation reduces to
\begin{equation}
\label{eq:phi3}
   \partial^\mu\partial_\mu\phi-\frac{1}{N^2-1}\left(1-\frac{1}{\xi}\right)(\eta^a\cdot\partial)^2\phi+Ng^2\phi^3=-j_\phi.
\end{equation}
where we have set $j_\phi=j/(N^2-1)$. Reducing to the case of the Lorenz gauge ($\xi=1$) being this equivalent to the Landau gauge at the classical level we are working on, this equation becomes classically exact (otherwise our solutions will be just asymptotic ones \cite{Frasca:2009yp}). The solution of this equation has been extensively discussed (\cite{Frasca:2010ce} and Refs. therein) and admits an iterative solution having the form at the leading order
\begin{equation}
   \phi(x)\approx \int d^4x' D(x-x')j_\phi(x')
\end{equation}
with
\begin{equation}
\label{eq:gluonD}
    D(p^2)=\sum_{n=0}^\infty\frac{Z_n}{p^2-m_n^2+i\epsilon}
\end{equation}
and
\begin{equation}
\label{eq:Zn}
    Z_n=(2n+1)\frac{\pi^2}{K^2(i)}\frac{(-1)^{n+1}e^{-(n+\frac{1}{2})\pi}}{1+e^{-(2n+1)\pi}}.
\end{equation}
being $K(i)=\int_0^{\frac{\pi}{2}}\frac{d\theta}{\sqrt{1+\sin^2\theta}}\approx 1.3111028777$, and a formula for the spectrum of the theory, in the strong coupling limit, given by
\begin{equation}
\label{eq:mn}
    m_n = \left(2n+1\right)\frac{\pi}{2K(i)}\left(\frac{Ng^2}{2}\right)^{\frac{1}{4}}\Lambda.
\end{equation}
From this ``mass spectrum'' we can identify a string tension when we set
\begin{equation}
    \sqrt{\sigma}=\left(\frac{Ng^2}{2}\right)^{\frac{1}{4}}\Lambda=(2\pi N\alpha_s)^{\frac{1}{4}}\Lambda.
\end{equation}
Here $\Lambda$ is just an arbitrary parameter arising from the integration of the equations of motion of the classical theory. It is not difficult to see, using Callan-Symanzik equation \cite{Frasca:2008gi}, that this propagator implies $\beta(Ng^2)=4Ng^2$ and so the theory is infrared free, making this argument completely consistent with a Gaussian generating functional.

So, there exists a class of classical solutions for the Yang-Mills equation in the Landau gauge that are consistent with the lattice results of a trivial infrared fixed point. We note the essential point that the sum of Yukawa propagators we obtain in eq.(\ref{eq:gluonD}) are exponentially damped and so, a few terms are already enough to fit a lattice propagator in the deep infrared limit. Our aim will be to compare it with a RGZ scenario and show that we are describing the same physics. This will provide fixed values to the free parameters, being these the condensates, of the RGZ scenario.

\section{Comparison with RGZ scenario}
\label{sec2}

From eq.(\ref{eq:Zn}) we can immediately evaluate the weight of each propagator into the series (\ref{eq:gluonD}). It is easy to see that
\begin{equation}
   Z_0\approx 1.14 \qquad Z_1\approx -0.15 \qquad Z_2\approx 0.01 \qquad  Z_3\approx -6.7\cdot 10^{-4} \ldots
\end{equation}
and so, taking the first few terms is enough to get a proper approximation in the deep infrared regime. Firstly, we fix the parameters for the case (\ref{eq:Drho}). We will have from eq.(\ref{eq:gluonD})
\begin{equation}
   D(p^2)\approx \frac{p^4 +p^2 (m_1^2Z_0+m_2^2Z_0+m_0^2Z_1+m_2^2Z_1+m_0^2Z_2+m_1^2 Z_2)
   +m_1^2 m_2^2 Z_0+m_0^2m_2^2Z_1+m_0^2m_1^2Z_2}{p^6+(m_0^2+m_1^2+m_2^2)p^4+(m_0^2m_1^2+
   m_0^2 m_2^2+m_1^2 m_2^2)p^2+m_0^2 m_1^2 m_2^2}
\end{equation}
where we have extracted an overall constant $C=Z_0+Z_1+Z_2\approx 1$. This gives, by a direct comparison with eq.(\ref{eq:Drho}),
\begin{eqnarray}
    2 M^2 &=& m_1^2Z_0+m_2^2Z_0+m_0^2Z_1+m_2^2Z_1+m_0^2Z_2+m_1^2 Z_2 \nonumber \\
    M^4 - \rho \rho^\dagger&=&m_1^2 m_2^2 Z_0+m_0^2m_2^2Z_1+m_0^2m_1^2Z_2 \nonumber \\
    m^2 + 2 M^2&=&m_0^2+m_1^2+m_2^2 \nonumber \\
    2 m^2 M^2 + M^4 + \lambda^4 - \rho \rho^\dagger&=&m_0^2m_1^2+m_0^2 m_2^2+m_1^2 m_2^2 \nonumber \\
    m^2 \left( M^4 - \rho \rho^\dagger \right)
                  + M^2 \lambda^4 - \frac{\lambda^4}{2}
                  \left( \rho + \rho^\dagger \right)&=&m_0^2 m_1^2 m_2^2.
\end{eqnarray}
In order to evaluate these parameters we have to fix the masses $m_n$. As chosen in Ref.\cite{Cucchieri:2012cb}, we take $\sqrt{\sigma}=0.44\ GeV$ for the string tension. This will give, using eq.(\ref{eq:mn}), $m_0=0.527\ GeV$, $m_1=1.58\ GeV$ and $m_2=2.63\ GeV$. We get
\begin{equation}
   M\approx 2.2\ GeV \qquad \rho \rho^\dagger\approx 4\ GeV^2 \qquad m^2=0.0033\ GeV^2
\end{equation}
and, already at this stage, we note that, using the following formula for the condensate \cite{Dudal:2010tf}
\begin{equation}
    \braket{g^2A^2}=-\frac{9}{13}\frac{N^2-1}{N}m^2
\end{equation}
one should have $m^2<0$. So, for consistency reasons we have to consider just the first couple of Yukawa propagators in eq.(\ref{eq:gluonD}) giving
\begin{equation}
    D(p^2)=(Z_0+Z_1)\frac{p^2+s}{p^4+u^2p^2+t^2}
\end{equation}
with
\begin{equation}
    s=M^2+\rho_1=\frac{Z_0m_1^2+Z_1m_0^2}{Z_0+Z_1} \qquad u^2=s+m^2=m_0^2+m_1^2 \qquad t^2=m^2s+\lambda^4=m_0^2m_1^2
\end{equation}
and in this case we get
\begin{equation}
    Z_0+Z_1\approx 0.99 \qquad s\approx 2.85\ GeV^2 \qquad u\approx 1.67\ GeV \qquad t\approx 0.83\ GeV^2
\end{equation}
that gives
\begin{equation}
    \braket{g^2A^2}=-\frac{9}{13}\frac{N^2-1}{N}m^2\approx 0.13\ GeV^2.
\end{equation}
%This result is in agreement with physical expectations in Ref.\cite{Narison:2002pw}. 
This gives a substantial validation to RGZ scenario for a gluon propagator in the form of eq.(\ref{prop}). Similarly, our scenario, as depicted in \cite{Frasca:2010ce} is perfectly consistent with RGZ scenario and provides further validation with respect to lattice data. Similarly, we have $\lambda^4=0.89\ GeV^4$. Finally, $D(0)=M^2/\lambda^4=5.44\ GeV^{-2}$ that is of the right magnitude as expected from lattice computations.

We note that, while the value of the other parameters is perfectly consistent with other determinations given in \cite{Dudal:2010tf,Cucchieri:2011ig}, the value of the dimension two condensate, $\braket{g^2A^2}$, is significantly smaller. The reason for this can be traced back to the dependence on the coupling and on the lattice spacing. In fact, the determination in \cite{Dudal:2010tf,Cucchieri:2011ig} is obtained from lattice data. Fitting at $\beta=6$ and $\beta=2.2$ provides quite different values for this condensate while we note that we are working at the trivial infrared fixed point however confirmed by RGZ scenario. This can be seen in a quite straightforward way. Using eq.(\ref{eq:mn}) for the mass spectrum we have
\begin{equation}
    m^2=m_0^2+m_1^2-s=m_0^2+m_1^2-\frac{Z_0m_1^2+Z_1m_0^2}{Z_0+Z_1}=-\frac{|Z_0+9Z_1|}{Z_0+Z_1}\frac{\pi^2}{4\sqrt{2}K^2(i)}\sqrt{N}g\Lambda^2
\end{equation}
and it is $m^2<0$ being $Z_1<0$ and $9|Z_1|>Z_0$. So, we note that, for lattice computations, are critical both coupling and lattice spacing and this explains the observed discrepancy in \cite{Dudal:2010tf,Cucchieri:2011ig}. Similarly, such a conclusion can also be drawn for $M^2+\rho_1$ and $\lambda^4$ and we have
\begin{equation}
   M^2+\rho_1=\frac{9Z_0+Z_1}{Z_0+Z_1}\frac{\pi^2}{4\sqrt{2}K^2(i)}\sqrt{N}g\Lambda^2 \qquad \lambda^4=\left[9+
   \frac{|Z_0+9Z_1|}{Z_0+Z_1}\left(10+\frac{|Z_0+9Z_1|}{Z_0+Z_1}\right)\right]\frac{\pi^4}{32K^4(i)}Ng^2\Lambda^4.
\end{equation}
This gives a proof of existence from first principles of the condensates in the RGZ scenario and, as an aside, we were able to fix them.

\section{Conclusions}
\label{sec3}
%%%%%%%%%%%%%%%%
%%%%%%%%%%%%%%%%%%%%%%%%%%%

In this paper we were able to show how RGZ scenario is perfectly consistent with a set of instanton solutions of Yang-Mills equations in the Landau gauge. The values of the condensates are finite and non-null making all the scenario well sound and explaining the excellent agreement with data on lattice. As an aside, we were able to fix in a closed form the condensates.

It is interesting to note that one of the first principles we started from to derive the condensates was the emerging data from lattice proving the existence of a trivial infrared fixed point for a pure Yang-Mills theory. This gives a Gaussian generating functional with the propagator having all the parameters fixed by the theory. With a proper set of Wightman axioms, it is possible to get a proper extension of the K\"allen-Lehman representation also for a gauge theory and this proves that the propagator one gets from the instanton solutions we obtain is indeed the one of a free theory. RGZ scenario also shows that the gluon propagator is the sum of Yukawa propagators and so both views validate each other.

As a final consideration, it is interesting to note that, with all the parameter properly fixed, RGZ scenario can be used to work out a lot of phenomenology. A gluonic Yukawa potential is enough to produce a Nambu-Jona-Lasinio model from QCD lagrangian.

%\section*{Acknowledgements}

%\newpage

% THANK YOU VERY MUCH CHARLES!!!
%\begin{acknowledgments}
%\end{acknowledgments}

%\bibliography{pra001}

\end{document}